\documentclass{easychair}

\usepackage{doc}
\usepackage{listings}
\usepackage[colorinlistoftodos]{todonotes}
\usepackage{url}
\usepackage{xcolor}
\usepackage{xspace}

\definecolor{burgundy}{rgb}{.909,.125,.467}
\definecolor{dkgreen}{rgb}{0,0.6,0}
\definecolor{gray}{rgb}{0.5,0.5,0.5}
\definecolor{mauve}{rgb}{0.58,0,0.82}
\definecolor{myblue}{rgb}{0.36,.34,.42}

\newboolean{showcomments}
\setboolean{showcomments}{true}

\newcommand{\inlinesection}[1]{\textbf{#1.}\xspace}

\newcommand{\abssolver}{\texttt{AbsSmtSolver}\xspace}
\newcommand{\abssort}{\texttt{AbsSort}\xspace}
\newcommand{\absterm}{\texttt{AbsTerm}\xspace}
\newcommand{\arraysort}{\texttt{ARRAY}\xspace}
\newcommand{\bool}{\texttt{BOOL}\xspace}
\newcommand{\bv}{\texttt{BV}\xspace}
\newcommand{\enum}{\texttt{enum}\xspace}
\newcommand{\function}{\texttt{FUNCTION}\xspace}
\newcommand{\intsort}{\texttt{INT}\xspace}
\newcommand{\ops}{\texttt{Ops}\xspace}
\newcommand{\op}{\texttt{Op}\xspace}

\newcommand{\real}{\texttt{REAL}\xspace}
\newcommand{\result}{\texttt{Result}\xspace}
\newcommand{\sat}{\texttt{SAT}\xspace}
\newcommand{\smtswitch}{\texttt{Smt-Switch}\xspace}
\newcommand{\solvers}{\texttt{SmtSolvers}\xspace}
\newcommand{\solver}{\texttt{SmtSolver}\xspace}
\newcommand{\sortkinds}{\texttt{SortKinds}\xspace}
\newcommand{\sortkind}{\texttt{SortKind}\xspace}
\newcommand{\sorts}{\texttt{Sorts}\xspace}
\newcommand{\sort}{\texttt{Sort}\xspace}
\newcommand{\termiterbase}{\texttt{TermIterBase}\xspace}
\newcommand{\termiter}{\texttt{TermIter}\xspace}
\newcommand{\terms}{\texttt{Terms}\xspace}
\newcommand{\term}{\texttt{Term}\xspace}
\newcommand{\uninterpreted}{\texttt{UNINTERPRETED}\xspace}
\newcommand{\unknown}{\texttt{UNKNOWN}\xspace}
\newcommand{\unsat}{\texttt{UNSAT}\xspace}
\newcommand{\virtual}{\texttt{virtual}\xspace}

\title{\smtswitch: a solver-agnostic C++ API for SMT solving\\
       (Extended Abstract)}

\author{Makai Mann\inst{1} \and
        Amalee Wilson\inst{1} \and
        Cesare Tinelli\inst{2} \and
        Clark Barrett\inst{1}}

\institute{
  Stanford University\\
  Stanford, USA
  \and
  The University of Iowa\\
  Iowa City, USA
 }

\authorrunning{Mann et al.}

\titlerunning{Smt-Switch}

\begin{document}

\maketitle

\begin{abstract}
  This extended abstract describes work in progress on \smtswitch, an
  open-source, solver-agnostic API for SMT solving. \smtswitch provides an
  abstract interface, which can be implemented by different SMT solvers.
  \smtswitch provides simple, uniform, and high-performance access to SMT
  solving for applications in areas such as automated reasoning, planning, and
  formal verification. The interface allows the user to create, traverse, and
  manipulate terms, as well as to dynamically dispatch queries to different
  underlying SMT solvers.
\end{abstract}

\section{Introduction}
\smtswitch is an open-source, solver-agnostic API for interacting with various
SMT solvers in C++. While SMT-LIB~\cite{smt-lib} provides a standard textual
interface for SMT solving, there are limitations. In particular, applications
that interact closely with a solver or its expressions could benefit from an
API. For example, tools that perform specialized term rewriting or dynamically
adjust queries based on satisfiability results. The two most common approaches
used by such tools now is to either build the tool on a specific SMT solver's
API, or write a custom expression representation which is then translated to
SMT-LIB and communicated to arbitrary SMT solvers through pipes as needed.
\smtswitch provides a convenient, in-memory representation that hides the
underlying SMT solver from the user. This allows projects to easily swap out
underlying SMT solvers, without developing a custom expression representation.

There already exists a well-established, solver-agnostic Python API:
PySMT~\cite{pysmt}. While this tool is excellent for prototyping, many
performant tools are not written in Python due to the overhead of an
interpreted language. Furthermore, PySMT keeps its own representation of terms
which is translated to underlying SMT solvers on demand. While this is extremely
convenient, it adds memory overhead and makes adding new theories more involved.
\smtswitch is a light wrapper on the underlying SMT solvers. This helps keep the
memory overhead low, and the maintenance of the tool straightforward. Adding support for
a new SMT solver to \smtswitch requires implementing an abstract interface using
the new solver's C or C++ API. The current version has implementations for
Boolector\cite{btor3}, CVC4\cite{cvc4}, MathSAT\cite{mathsat5}, and
Yices2\cite{yices2}. It supports quantifier-free formulas over booleans,
bitvectors, integers, reals, uninterpreted functions and arrays. The API can
easily be extended with new sorts and features supported by the underlying
solvers. The tool is available at \url{https://github.com/makaimann/smt-switch}.

\section{Design}
\smtswitch is designed to be a lightweight wrapper on the C or C++ API of the
underlying SMT solvers. It intentionally delegates as much of the functionality
to the underlying solvers as possible. This reduces redundancy, and results in
simpler implementations and lower memory overhead. The API is implemented in
C++11 and also provides Python bindings using Cython~\cite{cython}. The Python
bindings can be used directly in a Python project, but the intended usage is for
tools that use \smtswitch in C++, but want to expose \smtswitch functionality in
Cython-generated Python bindings for that tool.

\inlinesection{Architecture} \smtswitch provides abstract classes which must be
implemented by a wrapper for each underlying solver. The implementations satisfy
the interface requirements of the abstract classes and wrap the relevant objects
needed for that underlying solver's own API. Throughout this abstract, we use
\emph{underlying} solver to refer to the SMT solver we are wrapping and
\emph{backend} to refer to the \smtswitch wrapper. Any solver with a C or C++
API can be added to \smtswitch. At the \smtswitch API level, the user interacts
with smart pointers to the abstract classes. The \virtual method functionality
of C++ allows the pointer to dynamically call the relevant method of the derived
class (the backend) that is pointed to. This architecture allows the interface
to be agnostic to the underlying solver. The three primary abstract classes are:
(i) \abssort; (ii) \absterm; and (iii) \abssolver. The interface and method
names are based on SMT-LIB version 2~\cite{smt-lib}. Figure
\ref{abssolver-interface} shows a representative selection of the \abssolver
header file which details the abstract interface. Many methods were removed for
space. The \op class is not abstract and does not need to be implemented by the
backend. However, the backend's \abssolver implementation must interpret an \op
when building terms.

\begin{figure}
\begin{lstlisting}[
  language=C++,
  numbers=left,
  stepnumber=1,
  numberstyle=\color{gray},
  basicstyle=\ttfamily,
  keywords={return,uint64_t,string,virtual,const,void},
  keywordstyle=\color{blue}\ttfamily,
  stringstyle=\color{mauve}\ttfamily,
  morecomment={[s][\color{mauve}]{<}{>}},
  commentstyle=\color{dkgreen}\ttfamily,
  classoffset=2,
  morekeywords={TermVec, Term, Sort, SmtSolver, SortKind, SortVec, Op, Result},
  keywordstyle=\color{burgundy}\ttfamily\bfseries,
  ]
/** Abstract solver class to be implemented by each supported solver. */
class AbsSmtSolver
{
 public:
  /* Add an assertion to the solver
   * SMTLIB: (assert <t>)
   * @param t a boolean term to assert */
  virtual void assert_formula(const Term & t) = 0;
  /* Check satisfiability of the current assertions
   * SMTLIB: (check-sat)
   * @return a result object - see result.h */
  virtual Result check_sat() = 0;
  /* Get the value of a term after check_sat returns a satisfiable result
   * SMTLIB: (get-value (<t>))
   * @param t the term to get the value of
   * @return a value term */
  virtual Term get_value(const Term & t) const = 0;
  /* Create a sort
   * @param sk the SortKind (BOOL, INT, REAL)
   * @return a Sort object */
  virtual Sort make_sort(const SortKind sk) const = 0;
  /* Create a sort
   * @param sk the SortKind (BV)
   * @param size (e.g. bitvector width for BV SortKind)
   * @return a Sort object */
  virtual Sort make_sort(const SortKind sk, uint64_t size) const = 0;
  /* Create a sort
   * @param sk the SortKind (FUNCTION)
   * @param sorts a vector of sorts (last sort is return type)
   * @return a Sort object
   * Note: This is the only way to make a function sort */
  virtual Sort make_sort(const SortKind sk, const SortVec & sorts) const = 0;
  /* Make a boolean value term
   * @param b boolean value
   * @return a value term with Sort BOOL and value b */
  virtual Term make_term(bool b) const = 0;
  /* Make a symbolic constant or function term
   * SMTLIB: (declare-fun <name> (s1 ... sn) s) where sort = s1x...xsn -> s
   * @param name the name of constant or function
   * @param sort the sort of this constant or function
   * @return the symbolic constant or function term */
  virtual Term make_symbol(const std::string name, const Sort & sort) = 0;
  /* Make a new term
   * @param op the operator to use
   * @param terms vector of children
   * @return the created term */
  virtual Term make_term(const Op op, const TermVec & terms) const = 0;
};
\end{lstlisting}
\caption{Representative set of methods in the \abssolver interface.\label{abssolver-interface}}
\end{figure}

\inlinesection{Building and Linking} \smtswitch uses CMake~\cite{cmake}. The
build infrastructure is designed to be modular with respect to backend solvers.
By default, the configuration script does not add any solvers. \smtswitch will
build but has no functionality. To add a solver, the user must first obtain the
underlying solver's library. \smtswitch provides convenience scripts for
obtaining several of the underlying solvers and instructions for the rest. Then,
a command line flag to the \smtswitch configuration script will add the solver
to the build. Each solver backend results in a separate library, e.g. on Linux
there will be a \texttt{libsmt-switch.so} file as well as a
\texttt{libsmt-switch-cvc4.so} file if configured with \texttt{--cvc4}. This
allows the user to build \smtswitch once, but only link solver backends to their
project as needed. The configuration script also has options to enable static
and debug builds.

\inlinesection{Testing} We use \emph{GoogleTest}~\cite{googletest} for the C++ test
infrastructure and \emph{Pytest}~\cite{pytest} for the Python test infrastructure.
Tests are parameterized by solver so that one test can easily be run over all
solvers.

\inlinesection{Undefined Behavior} Each backend implementation of a solver needs
to recover its relevant objects from a generic abstract object. This is done
with a \texttt{static\_pointer\_cast}. This results in \emph{undefined behavior}
if the cast is not valid. In particular, this means that sharing \sorts and
\terms between different \solvers results in undefined behavior. To move a \sort
or \term between solvers, it must explicitly be transferred. There is a class
provided in the API which can transfer terms between solvers. \smtswitch is
intentionally lightweight and thus does not perform much error checking. This
design decision was made to reduce overhead and redundancy, since most SMT
solvers do error checking already. However, the class hierarchy allows
users to extend classes or build wrappers if they would like to insert their own
error checking at the \smtswitch level. \\ 

\inlinesection{Custom Exceptions} \smtswitch defines its own set of exceptions
that inherit from \texttt{std::exception}. Each of them take a
\texttt{std::string} message for describing the problem that can be accessed
with the \texttt{what} method.
\begin{enumerate}
  \item \texttt{SmtException} : the generic \smtswitch exception - all other
    custom exceptions inherit from it.
  \item \texttt{NotImplementedException} : the exception for an unimplemented
    feature in a backend solver.
  \item \texttt{IncorrectUsageException} : an exception that is thrown when
    incorrect usage on the users' part is detected.
  \item \texttt{InternalSolverException} : an exception that is thrown when
    there is an error in the underlying solver.
\end{enumerate}

\section{Abstract Interface}
\begin{figure}[t]
  \includegraphics[width=\textwidth]{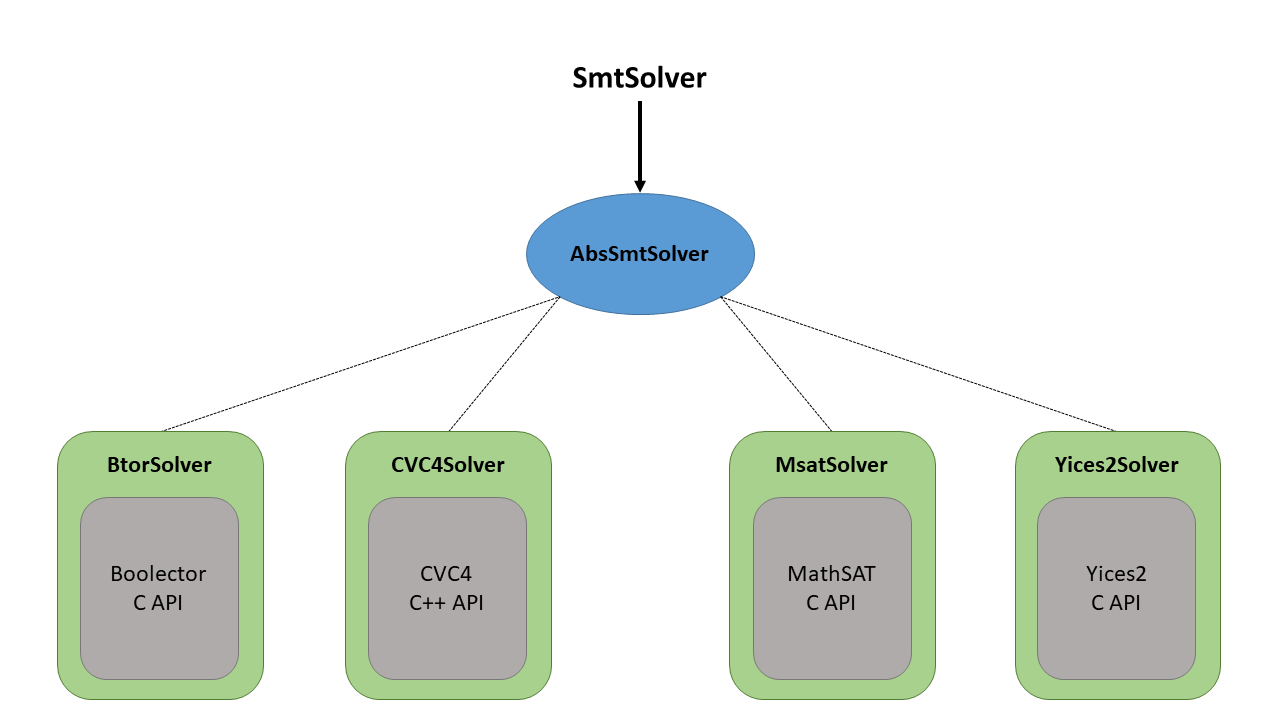}
  \caption{\abssolver Hierarchy. \solver is a \texttt{typedef} for a pointer to an
    \abssolver. The dashed lines represent inheritance.\label{fig:abssolver}}
  \vspace{-1em}
\end{figure}

We now describe the abstract interface in more detail. We start by describing
the \abssort class which illustrates the supported theories. This is followed by
the non-abstract struct for representating operators: \op. Next, we describe the
other two important abstract classes: \absterm and \abssolver. Figure
\ref{fig:abssolver} depicts the class hierarchy for an \abssolver. The \abssort
and \absterm classes have analogous architectures. Finally, we describe the
\result struct which is returned after a satisfiability query, and the solver
factories that are used to instantiate solvers. 

\subsection{AbsSort}
The \abssort abstract class represents the type of \terms created in \smtswitch.
Sorts are characterized by an \enum, \sortkind, which categorizes it and
additional parameters based on the \sortkind. \abssort has \virtual methods for
querying the sort for its \sortkind and parameters. This abstract class is
implemented by each backend solver and wraps the backend's representation of a
sort. For example, the CVC4 backend has a \texttt{CVC4Sort} class which wraps a
sort object from CVC4's C++ API. A \sort is a pointer to an \abssort. Currently
supported \sortkinds and associated parameters are the following:

\begin{enumerate}
\item \bool : booleans
\item \intsort : integer numbers
\item \real : real numbers
\item \bv : fixed-width bitvectors
  \begin{enumerate}
  \item parameter: positive integer width
  \end{enumerate}
\item \function : uninterpreted function sort
  \begin{enumerate}
  \item parameter: vector of domain \sorts
  \item parameter: the codomain \sort
  \end{enumerate}
\item \arraysort : arrays parameterized by index and element sorts
  \begin{enumerate}
    \item parameter : index \sort
    \item parameter : element \sort
   \end{enumerate}
\item \uninterpreted : an uninterpreted sort
  \begin{enumerate}
  \item parameter: non-negative integer arity
  \end{enumerate}
\end{enumerate}

\subsection{Op}
\op is a struct used to represent builtin operators for constructing terms in
\smtswitch. These can be split into two categories: primitive operators and
indexed operators. For unity of representation, an \op is used for both. An \op
stores a \texttt{PrimOp} \texttt{enum} and up to two integer indices. Primitive
operators are defined only by the \texttt{PrimOp} and take no indices. Indexed
operators have both a \texttt{PrimOp} and one or two indices. For convenience,
the \op constructor can be applied implicitly by the compiler. Thus, building
terms with a primitive operator can be accomplished by using a \texttt{PrimOp}
directly, without explicitly building an \op from it. \smtswitch uses a
straightforward naming scheme such that the \texttt{PrimOp} enums have
one-to-one correspondence with SMT-LIB operators.

\ops can be null. All \terms that are a symbol or value have a null operator.
Note that terms with null operators are not necessarily leaf nodes. For
example, one such edge case is constant arrays (arrays where every element has
been set to a given value). Constant arrays are themselves values and thus have
a null operator. However, they still have one child which is the value assigned
to every element.

\subsection{AbsTerm}
The \absterm abstract class represents expressions built through the API.
Constructing the term uses methods from the \solver because most underlying
solvers use a solver or context object for registering terms. However, once the
term is created, the \smtswitch interface assumes that it can be queried for
information. For example, a term has \virtual methods for accessing the \sort,
the \op used to create it, and the children of the \term. Querying the children
is implemented as an iterator. Thus, \terms have a \texttt{begin} and
\texttt{end} method that each return a \termiter, which is a class that wraps an
abstract class, \termiterbase. Each backend solver must implement a derived
class for \absterm that wraps its relevant expression representation(s). If the
backend solver supports accessing the children of a term, it must also implement
a derived class for \termiterbase. For example, the CVC4 backend has both a
\texttt{CVC4Term} and \texttt{CVC4TermIter} object that implement the \term and
\termiterbase interfaces and store the relevant term and term iteration objects
from the CVC4 C++ API. A \term is a pointer to an \absterm. \smtswitch also has
type declarations for convenient data structures such as \texttt{TermVec} which
is a vector of \terms.

A given \term can be a symbol (uninterpreted constant or function), a value
(theory values such as the number 1), or an expression built from symbols and
values using operators. \emph{All} SMT expressions in \smtswitch are represented
as terms. In particular, we take a higher-order logic perspective and represent
uninterpreted functions as terms. Thus, applying an uninterpreted function uses
an ``\texttt{Apply}'' operator on the function and the arguments.
Some SMT solvers are starting to add higher-order logic features, and this
representation is often convenient. For example, an invariant maintained by
\smtswitch is that any term with a non-null operator can always be rebuilt using
its operator and children. For a given \term t with a non-null operator, the
backend solver implementation should guarantee that the following always returns
\texttt{true}:
\begin{equation*}
\texttt{t == solver->make\_term(t->get\_op(), TermVec(t->begin(),
  t->end()))}
\end{equation*}
If the API instead had different methods for applying an
uninterpreted function versus creating terms with a builtin operator, this
invariant would not be maintained.

\subsection{AbsSmtSolver}
\abssolver declares the main interface that a user interacts with. It has
methods for declaring \sorts, building \terms, asserting formulas and checking
for satisfiability. \solver is a pointer to an \abssolver. The bulk of the
implementation for a backend solver will likely be its implementation of
\abssolver. The interface methods closely match the commands of SMT-LIB. The
naming scheme simply replaces ``-'' with ``\_''. For example, some methods of
\abssolver include \texttt{set\_opt}, \texttt{check\_sat}, and
\texttt{get\_value}. One notable exception is that assertions are added with the
\texttt{assert\_formula} method (as opposed to ``\texttt{assert}'' as in
SMT-LIB), to avoid clashing with the C \texttt{assert} macro.

Each backend solver must implement a derived class for \abssolver. For example,
the CVC4 backend has a \texttt{CVC4Solver} object that inherits \abssolver. This
derived class must store the relevant information for that solver and implement
each of the \virtual \abssolver methods in the underlying solver's API. The
\abssolver methods all operate over pointers to the abstract objects as shown in
Figure \ref{abssolver-interface}. For example, the method at line 47 has
signature: \texttt{make\_term(Op, const TermVec \&)}. The derived class
implementation must cast each of the \term pointers in the \texttt{TermVec} to
its own derived class implementation of \absterm to be able to access relevant
wrapped members. Thus, the \texttt{CVC4Solver} implementation of that method
would cast each of the \terms to a \texttt{CVC4Term} with
\texttt{static\_pointer\_cast<CVC4Term>(t)} for \term t. Then, it would intepret
the \op, perform the corresponding operation over the CVC4 objects, and finally
wrap the result in a \texttt{CVC4Term} and return it as a \term.

\subsection{Result}
\result is a struct for representing the return value of a call to
\texttt{check\_sat} or \texttt{check\_sat\_assuming}. It currently has three
possible values: \sat, \unsat, and \unknown. In the case of \unknown, the
backend solver can optionally provide a \texttt{std::string} with an explanation
of the reason for the unknown result.

\subsection{Solver Factories}
A solver factory is a simple class that defines a single static method:
\texttt{create(bool logging)}. Each backend solver implementation defines a
factory and has a dedicated header file. The \texttt{create} function is used to
create a \solver of that type. The single boolean parameter is to choose
whether or not the term DAG is tracked at the \smtswitch level.

If \texttt{logging} is set to \texttt{false}, the \solver relies on the
underlying solver API for querying a term for its \op, \sort and children (e.g.
term traversal) by translating back to \smtswitch objects. If \texttt{logging}
is set to \texttt{true}, the method returns the backend solver but wrapped by a
\texttt{LoggingSolver}. This is an implementation of \abssolver that forwards
commands to another backend \solver. Additionally, it keeps a term DAG at the
\smtswitch level. This is accomplished by wrapping every \sort and \term created
through the \texttt{LoggingSolver} in a \texttt{LoggingSort} and
\texttt{LoggingTerm}, respectively. In addition to storing the underlying
objects, these classes also keep relevant data for tracking the expression DAG
as it was constructed. For example, a \texttt{LoggingSort} keeps the \sortkind
and parameters used to create it. A \texttt{LoggingTerm} keeps the \op,
(Logging) \sort, and children (Logging) \terms. This optional feature can be
useful for underlying solvers that perform on-the-fly rewriting or alias sorts
(e.g., do not distinguish between bitvectors of width one and booleans). The
logging infrastructure ensures that a created term has exactly the same \op,
\sort, and children that were used to create it. The implicit assumption is that
even though creating a term through a solver API might result in a rewritten
term, creating the term again with the same \op and children will result in the
same rewritten term. Thus, the logging infrastructure hides the underlying
solver's rewriting from the user. To contend with sort aliasing, the
\texttt{LoggingSolver} also performs sort inference to compute the expected sort
when building terms. The logging infrastructure simplifies transferring terms
between different solvers (which might not alias sorts in the same way) and can
be more intuitive. Additionally, some solver backends (currently the Yices2
backend) rely on the logging infrastructure for term traversal. A Yices2 \solver
created without logging will create \terms that do not support iterating over
the children.

\section{Examples}
In this section we demonstrate the \smtswitch API with a simple example using
both the C++ API and the Python bindings. Figure \ref{cpp-example} uses
\smtswitch with the CVC4 backend to solve two simple queries over bitvectors and
uninterpreted functions. Lines 1-2 include the necessary \smtswitch headers, and
lines 3-8 are standard C++ includes, \texttt{using} declarations and main
function. Line 9 declares a \solver using CVC4 as the backend without logging.
In line 10 the solver is set to incremental mode. Lines 11 and 12 declare a
bitvector sort of width 9 and a function sort from a width 9 bitvector to a
width 9 bitvector. Lines 13-17 declare two bitvector symbolic constants, an
uninterpreted function and then applies the function to each of the bitvector
symbols. Line 20 uses a C assert to check that the operator of an applied
uninterpreted function term is \texttt{Apply}. Line 21 populates a vector with
the children of term \texttt{fx}. Line 22 asserts that there are two children
(in some solvers this term would have only one child: \texttt{x}). Lines 25 and
26 check that the children are the function, \texttt{f}, and the bitvector,
\texttt{x}. Line 29 asserts to the \solver that the returned values from the
function applied to x and y are different. Lines 30 and 31 extract the bottom 8
bits from \texttt{x} and \texttt{y}, respectively. Line 33 asserts to the
\solver that the bottom bits of \texttt{x} and \texttt{y} are equivalent. Line
35 checks the satisfiability of the current assertions and line 36 checks that
the solver found \sat as expected. The query is satisfiable because \texttt{x}
and \texttt{y} can have different most significant bits, and thus the function
applications could return different values. Line 38 asserts to the \solver that
the most significant bits of \texttt{x} and \texttt{y} are also equivalent.
Checking satisfiability in lines 41 and 42 confirms that the current set of
assertions are now unsatisfiable because of uninterpreted function axioms.
Figure \ref{python-example} shows exactly the same example but through the
Python API.

It is very simple to change the solver in these two examples, assuming that the
relevant solver libraries have already been built. In the C++ example this
amounts to including the relevant solver factory file (similar to line 2),
updating line 9 to use a different solver factory create function, and finally
recompiling and relinking with the new solver library. Assuming the Python
bindings were built with the relevant solver, changing the solver in the Python
example only requires using a different create function in line 5 of Figure
\ref{python-example}.

\begin{figure}
\begin{lstlisting}[
  language=C++,
  numbers=left,
  stepnumber=1,
  numberstyle=\color{gray},
  basicstyle=\ttfamily,
  keywords={return,false,true},
  keywordstyle=\color{blue}\ttfamily,
  stringstyle=\color{mauve}\ttfamily,
  morecomment={[s][\color{mauve}]{<}{>}},
  commentstyle=\color{dkgreen}\ttfamily,
  classoffset=2,
  morekeywords={TermVec, Term, Sort, SmtSolver},
  keywordstyle=\color{burgundy}\ttfamily\bfseries,
  ]
#include "smt-switch/smt.h"
#include "smt-switch/cvc4_factory.h"
#include "assert.h"
#include <iostream>
using namespace smt;
using namespace std;
int main()
{
  SmtSolver s = CVC4SolverFactory::create(false);
  s->set_opt("incremental", "true");
  Sort bvsort9 = s->make_sort(BV, 9);
  Sort funsort = s->make_sort(FUNCTION, {bvsort9, bvsort9});
  Term x = s->make_symbol("x", bvsort9);
  Term y = s->make_symbol("y", bvsort9);
  Term f = s->make_symbol("f", funsort);
  Term fx = s->make_term(Apply, f, x);
  Term fy = s->make_term(Apply, f, y);

  // Functions are terms
  assert(fx->get_op() == Apply);
  TermVec fx_children(fx->begin(), fx->end());
  assert(fx_children.size() == 2);
  // These equalities are structural e.g. the first child *is* f
  // These are not SMT equalities
  assert(fx_children[0] == f);
  assert(fx_children[1] == x);

  // (assert (distinct (f x) (f y)))
  s->assert_formula(s->make_term(Distinct, fx, fy));
  Term x_7_0 = s->make_term(Op(Extract, 7, 0), x);
  Term y_7_0 = s->make_term(Op(Extract, 7, 0), y);
  // (assert (= ((_ extract 7 0) x) ((_ extract 7 0) y)))
  s->assert_formula(s->make_term(Equal, x_7_0, y_7_0));

  Result r = s->check_sat();
  assert(r.is_sat()); // the MSB of x and y can be different
  // (assert (= ((_ extract 8 8) x) ((_ extract 8 8) y)))
  s->assert_formula(s->make_term(Equal,
                                 s->make_term(Op(Extract, 8, 8), x),
                                 s->make_term(Op(Extract, 8, 8), y)));
  r = s->check_sat();
  assert(r.is_unsat());
  return 0;
}
  \end{lstlisting}
  \caption{C++ API Example
  \label{cpp-example}}
  \end{figure}

\begin{figure}
\begin{lstlisting}[
    %frame=tb,
  language=Python,
  numbers=left,
  stepnumber=1,
  %aboveskip=3mm,
  %belowskip=3mm,
  showstringspaces=false,
  columns=flexible,
  basicstyle=\ttfamily,
  numberstyle=\color{gray},
  morekeywords={self, assert, False, True},
  keywordstyle=\bfseries\color{blue},
  deletendkeywords={eval},
  commentstyle=\color{dkgreen},
  stringstyle=\color{mauve},
  breaklines=true,
  breakatwhitespace=true,
  tabsize=3
  ]
import smt_switch as ss
from smt_switch.primops import Apply, Distinct, Equal, Extract

if __name__ == "__main__":
    s = ss.create_cvc4_solver(False)
    s.set_opt('incremental', 'true')
    bvsort9 = s.make_sort(ss.sortkinds.BV, 9)
    funsort = s.make_sort(ss.sortkinds.FUNCTION, [bvsort9, bvsort9])
    x = s.make_symbol('x', bvsort9)
    y = s.make_symbol('y', bvsort9)
    f = s.make_symbol('f', funsort)
    fx = s.make_term(Apply, f, x)
    fy = s.make_term(Apply, f, y)

    assert fx.get_op() == Apply
    fx_children = [c for c in fx]
    assert len(fx_children) == 2
    assert fx_children[0] == f
    assert fx_children[1] == x

    s.assert_formula(s.make_term(Distinct, fx, fy))
    x_7_0 = s.make_term(ss.Op(Extract, 7, 0), x)
    y_7_0 = s.make_term(ss.Op(Extract, 7, 0), y)
    s.assert_formula(s.make_term(Equal, x_7_0, y_7_0))

    r = s.check_sat()
    assert r.is_sat()
    s.assert_formula(s.make_term(Equal,
                                 s.make_term(ss.Op(Extract, 8, 8), x),
                                 s.make_term(ss.Op(Extract, 8, 8), y)))
    r = s.check_sat()
    assert r.is_unsat()
  \end{lstlisting}

  \caption{Python Bindings Example\label{python-example}}
  \vspace{-1em}
\end{figure}

\section{License}    

The \smtswitch code is distributed under the BSD 3-clause license. However, not
all solvers have BSD-compatible licenses. For these solvers, the user must
obtain the solver libraries themselves and ensure they meet all the requirements for
the license. There are then instructions for how to properly build smt-switch
with that solver as the backend, in which case the license assumes that of the
solver. The BSD-compatible backend solvers are Boolector and CVC4.

\section{Related Work}
The most closely related works are \texttt{smt-kit}~\cite{smt-kit} and
\emph{metaSMT}~\cite{metaSMT}, other C++ APIs for SMT solving. Both APIs utilize
templates to be solver agnostic, and have term representations that are separate
from the underlying solver, as opposed to \smtswitch which only provides an
abstract interface and a light wrapper around the underlying solvers. This
design choice reduces overhead and keeps maintenance simple. \emph{metaSMT}
makes clever use of C++ template meta-programming which allows compile-time
translation to the underlying SMT solver's API when possible, resulting in
extremely low overhead. Furthermore, it provides several features including
bit-blasting and infrastructure for portfolio solving. However, \emph{metaSMT}
is less focused on runtime decisions for expression building or queries, and
only supports bit-vectors, arrays, and uninterpreted functions. Adding new
theories to either \texttt{smt-kit} or \emph{metaSMT} would likely be a bigger
undertaking than in the comparatively simple \smtswitch. Neither
\texttt{smt-kit} nor \emph{metaSMT} appear to be under active development since
2014 and 2016, respectively.

Two other related tools are \texttt{PySMT}~\cite{pysmt} and
\texttt{sbv}~\cite{sbv}. \texttt{PySMT} is a solver-agnostic SMT solving API for
Python. \texttt{PySMT} has its own term representation and translates formulas
to the backend solvers dynamically once they are asserted. It also uses a class
hierarchy to support swapping underlying solvers. \texttt{sbv} is a
solver-agnostic SMT-based verification tool for Haskell. It provides its own
datatypes for representing various SMT queries and communicates with solvers
through SMT-LIB with pipes.

\section{Conclusion}
We have presented work in progress on \smtswitch, a solver-agnostic API for
performant prototyping with SMT solvers in C++. This system is available for use
on GitHub and has already been used in projects~\cite{pono,lazybv2int}. Future work includes
adding a backend for Z3~\cite{z3}, adding support for quantifiers and inductive
datatypes, as well as investigating possible performance improvements, such as
optimized data structures and alternatives to smart pointers.

\newpage

\bibliographystyle{plain}
\bibliography{ms}

\end{document}